\documentclass[aps, pra, superscriptaddress, twocolumn, 10pt, floatfix, nofootinbib, tightenlines]{revtex4-1}
\pdfoutput=1

\usepackage{blindtext}
\usepackage{times,bbm,amsmath,amssymb,braket}
\usepackage{soul}
\usepackage{xcolor}
\usepackage{bm}
\usepackage{graphicx}
\usepackage{appendix}
\usepackage[acronym]{glossaries}
\usepackage{hyperref}
\hypersetup{
    colorlinks=true,
    linkcolor=blue,
    filecolor=magenta,      
    urlcolor=cyan,
}

\newacronym{spdc}{SPDC}{Spontaneous Parametric Down Conversion}
\newacronym{mse}{MSE}{Mean Squared Error}
\newacronym{crb}{CRB}{Cramér-Rao Bound}

\begin{document}
\title{Experimental investigation of Bayesian bounds in multiparameter estimation}

\author{Simone Evaldo D'Aurelio}
\affiliation{Dipartimento di Fisica, Sapienza Universit\`a di Roma, P.le Aldo Moro 5, 00185, Rome, Italy}
\author{Mauro Valeri}
\affiliation{Dipartimento di Fisica, Sapienza Universit\`a di Roma, P.le Aldo Moro 5, 00185, Rome, Italy}
\author{Emanuele Polino}
\affiliation{Dipartimento di Fisica, Sapienza Universit\`a di Roma, P.le Aldo Moro 5, 00185, Rome, Italy}
\author{Valeria Cimini}
\affiliation{Dipartimento di Fisica, Sapienza Universit\`a di Roma, P.le Aldo Moro 5, 00185, Rome, Italy}

\author{Ilaria Gianani}
\affiliation{Dipartimento di Scienze,  Universit\`a degli Studi Roma Tre, Via della Vasca Navale 84, 00146, Rome, Italy}
\author{Marco Barbieri}
\affiliation{Dipartimento di Scienze,  Universit\`a degli Studi Roma Tre, Via della Vasca Navale 84, 00146, Rome, Italy}

\author{Giacomo Corrielli}
\affiliation{Istituto di Fotonica e Nanotecnologie - Consiglio Nazionale delle Ricerche (IFN-CNR), p.za Leonardo da Vinci 32, 20133 Milano}
\affiliation{Dipartimento di Fisica - Politecnico di Milano, p.za Leonardo da Vinci 32, 20133 Milano}
\author{Andrea Crespi}
\affiliation{Dipartimento di Fisica - Politecnico di Milano, p.za Leonardo da Vinci 32, 20133 Milano}
\affiliation{Istituto di Fotonica e Nanotecnologie - Consiglio Nazionale delle Ricerche (IFN-CNR), p.za Leonardo da Vinci 32, 20133 Milano}
\author{Roberto Osellame}
\affiliation{Istituto di Fotonica e Nanotecnologie - Consiglio Nazionale delle Ricerche (IFN-CNR), p.za Leonardo da Vinci 32, 20133 Milano}
\affiliation{Dipartimento di Fisica - Politecnico di Milano, p.za Leonardo da Vinci 32, 20133 Milano}

\author{Fabio Sciarrino}
\affiliation{Dipartimento di Fisica, Sapienza Universit\`a di Roma, P.le Aldo Moro 5, 00185, Rome, Italy}
\author{Nicol\`o Spagnolo}
\affiliation{Dipartimento di Fisica, Sapienza Universit\`a di Roma, P.le Aldo Moro 5, 00185, Rome, Italy}

\begin{abstract}
Quantum parameter estimation offers solid conceptual grounds for the design of sensors enjoying quantum advantage. This is realised not only by means of hardware supporting and exploiting quantum properties, but data analysis has its impact and relevance. In this respect, Bayesian methods have emerged as an effective and elegant method, with the perk of incorporating naturally the availability of {\it a priori} information. In this article we present an evaluation of Bayesian methods for multiple phase estimation, assessed based on bounds that work beyond the usual limit of large samples assumed in parameter estimation. Importantly, such methods are applied to experimental data generated from the output statistics of a three-arm interferometer seeded by single photons. Our studies provide a blueprint for a more comprehensive data analysis in quantum metrology. 
\end{abstract}

\maketitle 
\section{Introduction}
The aim of quantum metrology is to study how the use of quantum phenomena can be beneficial in parameter estimation \cite{giovannetti2006prl,paris2009ijqi,giovannetti2011natphot, escher2011noisy,toth2014review,braun2018review,pirandola2018review,polino2020review,sidhu2020review}. As the technological level of quantum sensors increases, ensuring reliable and efficient data processing becomes a crucial aspect in the operation of devices. While in proof-of-principle demonstrations the enhancement in precision can be shown with simple data analysis, for real applications optimal use of the available data is required \cite{gianani2020ieee}: this is instrumental to reach the necessary accuracy, as well as the ultimate limits on precision.

The estimation process relies on two sources of information. On the one hand, we may have some {\it a priori} information available on our parameter from modelling or preliminary measurements. On the other hand, the collected data allow to refine our previous knowledge. Bayesian data analysis thus provides the most natural setting to combine both contributions to deliver a high-quality estimator~\cite{helstrom1976quantum,box2011bayesian}. The standard approach to assess its performance in terms of precision is to compare the experimental variance with that of the Cram\'er-Rao bound~\cite{helstrom1976quantum}.
This is determined by the Fisher information, which indicates the amount of information embedded in the output probability of the quantum sensor. However, this approach is valid only for local estimation, {\it i.e.} in a small range around a known value of the parameter. This may not be the case in general, and suitable methods to achieve local conditions are needed~\cite{hentschel2010adaptive1,hentschel2011adaptive2,lovett2013differential,wiebe2016efficient,paesani2017bayesian,lumino2018swarm, rambhatla2020genetic,liu2021optimal,fiderer2021neural,nolan2021machine,craigie2021resource,han2021adaptive}. More exhaustive data processing thus has to incorporate cases with relatively broad {\it a priori} information. This is particularly relevant when limited metrological resources can be invested for the measurement \cite{rubio2018jphyscomm,rubio2019njp,rubio2020bayesian,PhysRevLett.124.030501,cimini2021non}. 

The interplay between {\it a priori} information and collected data becomes even more complex in the multiparameter case. Here, the estimation procedure is affected not only by the initial uncertainties on the individual parameters, but also on their correlations. Indeed, the multiparameter Fisher information sets a given degree of statistical correlation between parameters \cite{paris2009ijqi,liu2020quantum}, which may contrast with the one obtained from {\it a priori} considerations, especially in its orientation in the parameter space to be estimated.

Adequate bounds on the variance of Bayes estimators have been introduced by Ziv and Zakai (ZZ)~\cite{ziv1969some}, and by Van Trees (VT)~\cite{vantrees1968book}, but have never been tested on data from a quantum sensing experiment. In this article, we show the applicability of the ZZ and VT bounds in multiphase integrated photonic quantum sensing. This is emerging as one of the most solid technologies for the realisation of quantum sensors~\cite{humphreys2013quantum,ciampini2016quantum,polino2019optica, valeri2020npj}. Our results show that both bounds capture the limits in multiphase estimation, especially for limited resources. The VT approach, besides being less computationally demanding, provides a tighter bound on the experimental variance. The ZZ approach still provides a useful reference, and captures the dependence on the variance on the correlations present in the {\it a priori} information.  
Our findings contribute to shaping actual data analysis for future real-life quantum sensors.

\section{Bayesian multiparameter estimation with a priori knowledge}

Bayesian estimation theory~\cite{box2011bayesian,helstrom1976quantum,li2019bayes} represents a paradigm for parameter estimation protocols and has been extensively and fruitfully employed \cite{polino2020review, berry2015quantum,armen2002adaptive,wheatley2010adaptive,higgins2007entanglement,lumino2018swarm,berni2015ab,paesani2017bayesian,daryanoosh2018experimental,roccia2018multiparameter,dinani2019bayesian,cimini2019quantum,li2019bayes,valeri2020npj}. In general, in multiparameter problems one has to deal with a set of physical parameters $\boldsymbol{\phi} = (\phi_1, \ldots, \phi_n)$, whose (unknown) values have to be estimated by the user. Typically, information on such set $\boldsymbol{\phi}$ is obtained by preparing a set of $N$ probe states $\rho_0$ of an auxiliary system. The evolution of the auxiliary system depends on the set $\boldsymbol{\phi}$ via a unitary operation $U(\boldsymbol{\phi})$, or more generally via a physical map $\mathcal{L}_{\boldsymbol{\phi}}$. Measurement $\Pi_{x}$ of the $N$ probe states $\rho_{\boldsymbol{\phi}}$ after evolution provides information on the parameters $\boldsymbol{\phi}$. 

\begin{figure}[ht!]
\centering
\includegraphics[width = 0.49\textwidth]{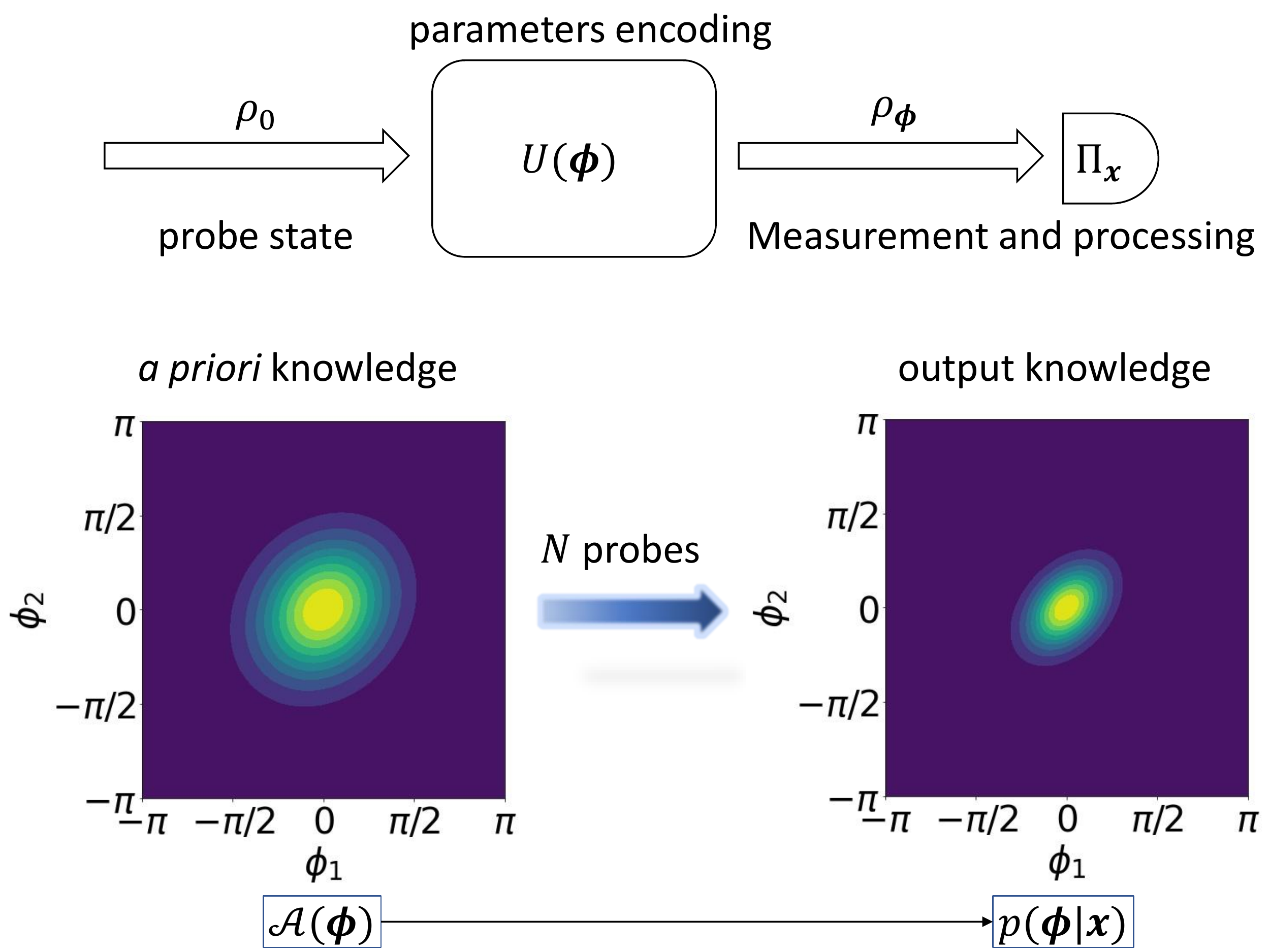}
\caption{General scheme for Bayesian parameter estimation. $N$ probes $\rho_0$ are sent through the system, acquiring information on the parameters $\boldsymbol{\phi}$, retrieved after measurement $\Pi_{x}$. The measurement outcomes $\boldsymbol{x}$ are used to update the initial knowledge $\mathcal{A}(\boldsymbol{\phi})$ to obtain the output conditional probability $p(\boldsymbol{\phi} \vert \boldsymbol{x})$, representing the result of the estimation process.}
\label{fig:fig1}
\end{figure}

Such process can be embedded in a Bayesian framework, where the parameters $\boldsymbol{\phi}$ to be estimated are treated as random variables distributed according to some distribution $p(\boldsymbol{\phi})$, which encodes the actual knowledge on the parameters values (see Fig. \ref{fig:fig1}). 
Bayesian estimation permits to naturally encode in such framework the availability of an arbitrary {\it a priori} information on the parameters, that may occur in several scenarios. This initial knowledge is encoded in the so-called prior distribution $\mathcal{A}(\boldsymbol{\phi})$, which acts as the starting point for the estimation process. After preparing $N$ probe states, the measurement outcomes $\boldsymbol{x} = (x_1, \ldots, x_N)$ are collected. Here, this vector represents the set of the $N$ measurements obtained by sending the sequence of inputs, where the outcome on a single probe $x_{k}=1, \ldots, s$ can take one of $s$ possible values according to the system behavior. Prior knowledge is updated according to the Bayes' rule: $p(\boldsymbol{\phi} \vert \boldsymbol{x}) \propto p(\boldsymbol{x} \vert \boldsymbol{\phi}) \mathcal{A}(\boldsymbol{\phi})$. Here $p(\boldsymbol{x} \vert \boldsymbol{\phi})$ is the conditional probability of detecting the sequence $\boldsymbol{x}$ of outcomes provided a given set of values $\boldsymbol{\phi}$ for the parameters. For $N$ independent probes, and individual measurements, such conditional probability is expressed as $p(\boldsymbol{x} \vert \boldsymbol{\phi}) = \prod_{k=1}^{N} p(x_k \vert \boldsymbol{\phi})$, where $p(x_{k} \vert \boldsymbol{\phi})$ is the likelihood of the given outcome $x_{k}$ for a single copy of the probe.

The output distribution $p(\boldsymbol{\phi} \vert \boldsymbol{x})$ fully encodes all the final knowledge on the parameters at the end of the estimation process. A common choice for an estimator of $\boldsymbol{\phi}$ is provided by its expectation value $\hat{\boldsymbol{\phi}}=\int \boldsymbol{\phi} p(\boldsymbol{\phi} \vert \boldsymbol{x}) \,d\boldsymbol{\phi}$, that can be proven to be an  asymptotically efficient and unbiased  estimator \cite{van2007bayesian,pezze2014,lehmann2006theory}. In single-parameter estimation, the precision can be quantified by the mean square error $\mathrm{MSE}(\phi)=\sum_{\boldsymbol{x}} p(\boldsymbol{x} \vert \phi) (\hat{\phi}-\phi)^2$. This generalizes in the multiparameter scenario to a matrix $C(\boldsymbol{\phi})$, with elements $C_{ij}(\boldsymbol{\phi}) = \sum_{\boldsymbol{x}} p(\boldsymbol{x} \vert \boldsymbol{\phi}) (\hat{\phi}_i - \phi_i) (\hat{\phi}_j - \phi_j)$. Such quantities substantially encode the precision in the estimate with respect to the true values of the parameters. Different theoretical studies focused on the extension of the Bayesian framework to the multiparameter case, also paying attention to the limited data regime \cite{gill2008conciliation,demkowicz2020multi,zhang2014quantum,lu2016quantum,rubio2020bayesian}.

In general, the mean square error (or its corresponding matrix form for a multiparameter scenario), is not accessible by the user, given the inherent absence of knowledge of the true parameter value. Bayesian approaches provide an additional framework to describe the estimation process. In particular, for a given experimental instance with $N$ probe states and the string $\boldsymbol{x}$ of measurement outcomes, the posterior distribution $p(\boldsymbol{\phi} \vert \boldsymbol{x})$ encodes the degree of confidence resulting from the estimation process via its covariance matrix $\Sigma$, whose elements are given by $\Sigma_{ij}=\int (\hat{\boldsymbol{\phi}}_i-\boldsymbol{\phi}_i)(\hat{\boldsymbol{\phi}}_j-\boldsymbol{\phi}_j) p(\boldsymbol{\phi} \vert \boldsymbol{x}) \,d\boldsymbol{\phi}$. The diagonal terms of $\Sigma$ represent the variances of the individual parameters. Conversely, off-diagonal terms contain information about the correlation between the parameters. In the limit of large $N$, given that Bayesian estimation is unbiased, it can be shown that the covariance matrix $\Sigma$ and matrix $C$ asymptotically coincide.

Different quantities can be used to bound the estimation error, as a function of the resources $N$ employed in the process. The most common bound applying to the covariance matrix is the celebrated Cram\'er-Rao bound~\cite{helstrom1976quantum} 
\begin{equation}
    \Sigma \geq \frac{1}{N} F(\boldsymbol{\phi})^{-1},
\end{equation}
where $N$ is the number of repetitions of the experiment, and the Fisher information matrix $F(\boldsymbol{\phi})$ has entries $F_{i,j}(\boldsymbol{\phi}) = \sum_{x} \partial_{\phi_i}p(x|\boldsymbol{\phi})\partial_{\phi_j}p(x|\boldsymbol{\phi})/p(x|\boldsymbol{\phi})$. This limit holds for local estimation: this means that the measurement assumes previous knowledge of the parameters $\boldsymbol \phi$, up to some uncertainty we aim at improving. Such an improvement may come from performing adaptive measurements, or, as it is often the case, by increasing the number of events $N$. Referring to the discussion above, in this large $N$ limit, the width of the posterior distribution has little to do with the {\it a priori} distribution $\mathcal{A}(\boldsymbol{\phi})$, and is governed uniquely by the Bayesian update.

When a limited number of repetitions are available, the Cram\'er-Rao limit may indeed not provide relevant information. However, if the prior distribution is regular and derivable, there exists a simple result due to Van Trees, which states that the matrix
\begin{equation}
\label{VTmatrix}
    H_{i,j} =  \int \mathcal{A}(\boldsymbol{\phi)}F_{i,j}(\boldsymbol{\phi})d\boldsymbol{\phi} +\frac{1}{N} \int \frac{\partial_{\phi_i}\mathcal{A}(\boldsymbol{\phi)}\partial_{\phi_j}\mathcal{A}(\boldsymbol{\phi)}}{\mathcal{A}(\boldsymbol{\phi)}}d\boldsymbol{\phi},
\end{equation}
sets a limit to the covariance as
\begin{equation}
\label{eq:VTB}
    \Sigma \geq \frac{1}{N} H^{-1},
\end{equation}
which is called Van Trees (VT) bound~\cite{gill1995applications}. This considers how both the average Fisher information available following the measurement and the prior enter in the final information; when $N$ is large, the usual Cram\'er-Rao bound is recovered, as the contribution from the {\it a priori} distribution becomes negligible.

For the VT bound to exist, $\mathcal{A}(\boldsymbol{\phi})$ must be regular; in order to obtain a valid bound also for generalised distributions, we adopt the Ziv-Zakai (ZZ) bound~\cite{ziv1969some}, in the multiparameter form proposed by Bell et al.~\cite{bell1997extended}. Differently from the VT bound, the ZZ bound is written explicitly in a scalar form, by introducing a unit vector $\bf{u}$ in the parameter space, and considering the error $\bf{u}^\top\Sigma\bf{u}$:
\begin{equation}
\label{eq:ZZB}
    \begin{gathered}
    \mathbf{u}^\top\mathbf{\Sigma}\mathbf{u}\geq\frac{1}{2}\int_0^\pi \,\tau\Bigg\{\max\limits_{\mathbf{v:u^Tv}=1}\int 
    [\mathcal{A}(\boldsymbol{\phi})+\mathcal{A}(\boldsymbol{\phi}+{\bf v}\tau)]\times \\P_e(\boldsymbol{\phi},\boldsymbol{\phi}+\mathbf{v}\tau)d\boldsymbol{\phi}\Bigg\}d\tau =  \mathcal{Z}(\bf u) 
    \end{gathered}
\end{equation}
We can define the error probability of distinguishing $\boldsymbol{\phi}$ from $\boldsymbol{\phi'}$ as:
\begin{equation}
    P_e(\boldsymbol{\phi},\boldsymbol{\phi'}) =\frac{1}{2}\left(1-\sum_x 
\vert \pi_0 p({\bf }x|\boldsymbol{\phi}) - \pi_1 p({\bf }x|\boldsymbol{\phi'})\vert^2\right),
\end{equation}
based on our measurements and the probabilities $\pi_0=1-\pi_1=\mathcal{A}(\boldsymbol{\phi})/(\mathcal{A}(\boldsymbol{\phi})+\mathcal{A}(\boldsymbol{\phi'}))$. These expressions hint at how the ZZ method adopts binary hypothesis testing as a way to bound the error, and avoids the need of a regular {\it a priori} distribution.

\section{Platform and methodology for Bayesian multiphase estimation}

\begin{figure}[ht!]
    \centering
    \includegraphics[width = 0.49\textwidth]{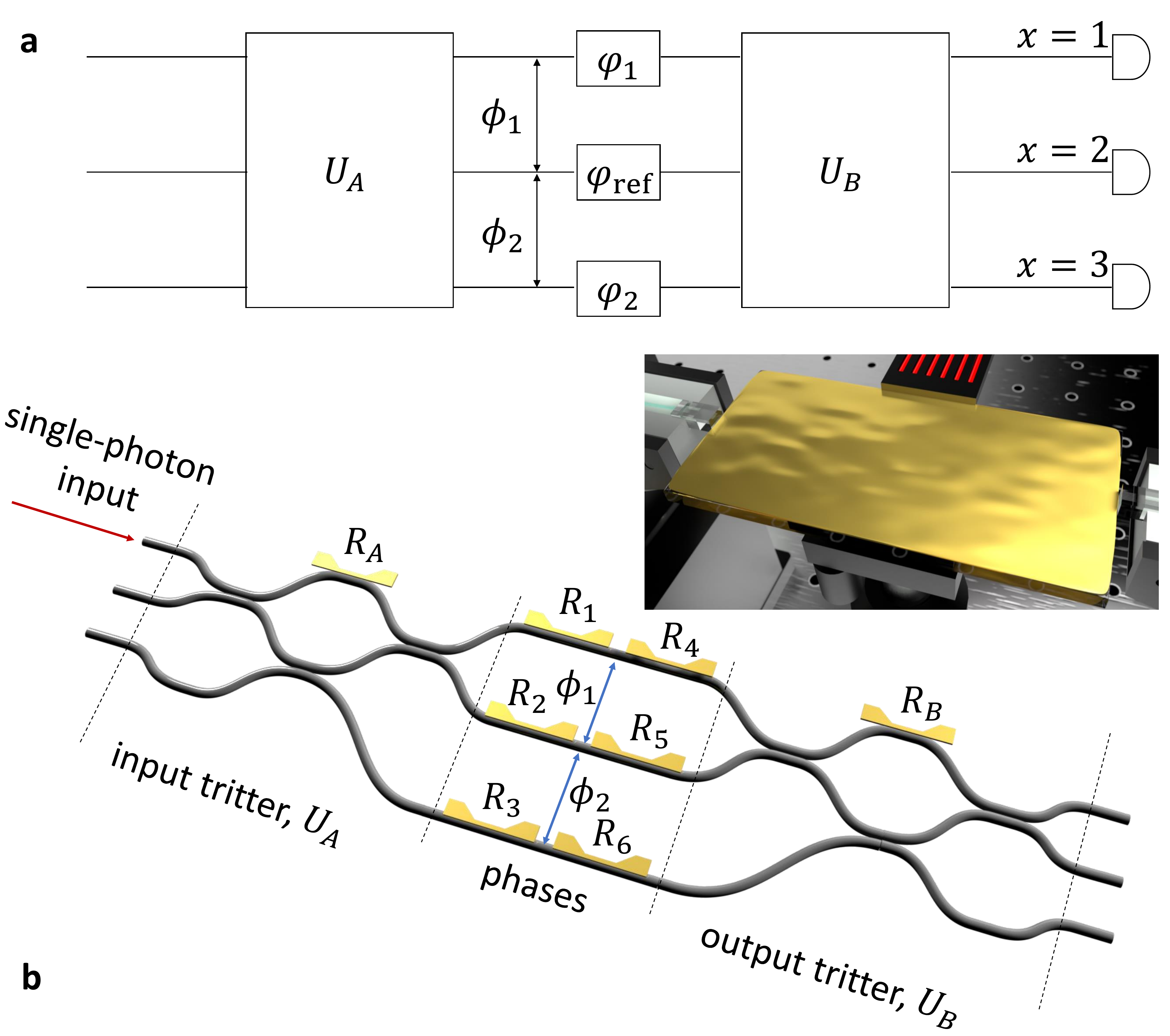}
    \caption{{\bf a}, Conceptual scheme of a multiarm interferometer for multiphase estimation. The relative phases to be estimated ($\phi_1 = \varphi_1 - \varphi_{\mathrm{ref}}, \ldots, \phi_{m-1} = \varphi_{m-1} - \varphi_{\mathrm{ref}}$) with respect to a common reference arm ($\varphi_{\mathrm{ref}}$) are embedded between input ($U_A$) and output ($U_B$) multimode transformations. Finally, single-photon detectors placed on the output modes provide the $m$ possible outcomes $x=1,\ldots,m$ for the measurement on each probe (figure shows the $m=3$ modes case). {\bf b}, Scheme of the reconfigurable interferometer employed for multiphase estimation. The first transformation $U_{A}$ is a balanced tritter, and includes a thermo-optic phase shifter placed on resistor $R_{A}$ used to adjust the transmittivities for balanced operations. The phases $\phi_1$ and $\phi_2$ to be estimated, which can be changed via the set of resistors $R_{1}-R_{6}$, are the two relative phases with respect to the central mode, acting as a reference. Output transformation $U_{B}$ is a second balanced tritter, which can be fine-tuned via resistor $R_{B}$, and it is used to recombine the output modes before the measurement.}
    \label{fig:interferometer}
\end{figure}

We have tested Bayesian estimation with {\it a priori} knowledge in an experimental platform provided by an integrated multiarm interferometer realized via the femtosecond laser writing technique \cite{Osellame2012} on a glass chip, where the unknown parameters are provided by two relative phase shifts inside the structure (see Fig. \ref{fig:interferometer}). In this way we apply the Bayesian framework and study the corresponding bounds  for the paradigmatic problem of multiphase estimation, having several applications to sensing and microscopy \cite{macchiavello2003optimal,humphreys2013quantum,polino2019optica,goldberg2020multiphase,gagatsos2016gaussian,gebhart2021bayesian,guo2019distributed,gessner2020multiparameter}.
More specifically, the interferometer is the three-mode generalization of a Mach-Zehnder one \cite{ciampini2016quantum,polino2019optica}. This is realized by replacing the input and output beam-splitters with two cascaded tunable tritters \cite{anton2019}. The output state of the interferometer will then depend on two different relative phase shifts between the internal modes. Within the circuit, phases can be actively reconfigured via thermo-optic phase shifters, obtained by fabricating resistive microheaters on top of the interferometer \cite{flamini2015lsa,polino2019optica,valeri2020npj}. In the present device, resistors $R_A$ and $R_B$ are used to fine-tune the input and output tritter transformations. The phase shifts for the estimation process are the two relative phases ($\phi_1, \phi_2$) referred to the central internal arm of the interferometer. These phases can be modified via the set of thermo-optic shifters corresponding to resistors $R_1-R_6$ and whose response functions can be also calibrated using machine learning techniques \cite{cimini2021calibration}.
Redundancy in such set of resistors can be used to implement adaptive estimation protocols, as shown in Ref. \cite{valeri2020npj}. Single-photon probes are conditionally generated from a two-photon source based on the spontaneous parametric down-conversion process. Namely, one of the two generated photon is directly measured, acting as a trigger for the experiments, while the other photon is injected in the integrated device via a single-mode fiber array in the first input port. After evolution through the interferometer, the output modes are coupled to an array of multimode fibers and measured via single-photon detectors.

As a first step, prior to the estimation experiments, single-photon states are employed to characterize the response of the interferometer. In particular, this amounts to experimentally reconstruct the correspondence between the current propagating on each resistor and the applied optical phase on the waveguide below, including potential cross-talks between the different tunable phase shifts. Furthermore, the same measurements are employed to reconstruct the transmittivity values for each directional coupler. This characterization is performed via the procedure reported in \cite{polino2019optica,valeri2020npj}.
More in details, single-photon input states are sent in the interferometer, and the output statistics for different values of the current are recorded. These data are then employed to perform a fitting procedure, using a mathematical model for the interferometer. Such a model includes both static parameters such as directional coupler transmittivities ($T_{j}$) and static phases obtained at zero current in the resistors ($\phi_{0j}$), and the dynamical response function $\phi_{l} = \phi_{l}(\{i_{j}, R_{j}\})$, being $R_{j}$ the resistances and $i_{j}$ the applied currents \cite{polino2019optica,valeri2020npj}. This procedure finally leads to knowledge of the likelihood function $P_{ML}(x \vert \phi_1, \phi_2)$.
Having access to such response function has a fundamental role for Bayesian estimation experiments, since this approach is based on progressively updating the posterior distribution according to the measurement outcomes and the likelihood function. Indeed, an accurate knowledge of the system likelihood is required to avoid biases or additional errors in phase estimation processes, and is a common need shared by quantum sensor platforms based on different technologies.

Then, phase estimation experiments are performed on the same system, characterizing the process for different values of the prior distribution $\mathcal{A}(\phi_1,\phi_2)$. The posterior distribution after each experiment with $N$ resources is updated from Bayesian update as described above, and carries all the available information on the unknown parameters. Relevant quantities, such as the estimate as well as the confidence interval related to the covariance matrix $\Sigma$, are extracted from this function. Bayesian protocols require evaluation of multidimensional integral, whose dimension is given by the number of unknown parameters. Such integrals become progressively more difficult to handle from a computational point of view. To address such issue, a possible solution is provided by performing an appropriate discretization of the parameters space, which can be performed by using the particle approximation \cite{granade2012njp}. Such a technique corresponds to replacing the continuous parameter space with a discrete set of $M^p$ points $\boldsymbol{y}_{i} = (\phi_{1,i}, \phi_{2,i})$, named particles. Each particle has its own weight $w_i$, related to the conditional probability associated with the corresponding values $\boldsymbol{y}_{i}$. More specifically, the initial particle set is obtained by randomly generating $M^p$ values $\boldsymbol{y}_{i}$ according to the prior distribution. The corresponding initial weights are uniformly set to $w_i = 1/M^p$. Bayesian update is then performed by changing at each step the weights of the particle according to the Bayes' rule. Thus, at each step $k$ the weights of the particles are updated as $w_i \rightarrow w_{i} P_{ML}(x_{k} \vert \phi_{1,i}, \phi_{i,2})$, followed by proper renormalization. After the $N$ steps of the estimation have been performed, the estimates are obtained as: 
\begin{equation}
\label{eq:est_particles}
    \hat{\boldsymbol{\phi}} = \sum_{i=1}^{M^p} w_i \boldsymbol{y}_{i},
\end{equation}
while the covariance matrix can be estimated as:
\begin{equation}
\label{eq:sigma_particles}
    \Sigma=\sum_{i=1}^{M^p} w_i (\mathbf{y}_i-\hat{\boldsymbol{\phi}}).(\mathbf{y}_i-\hat{\boldsymbol{\phi}})^T.
\end{equation}
In our case, given the intermediate $N$ regime and the intrinsic periodicity of the phase parameters, we employed the corresponding circular counterparts for the estimate and the covariance matrix \cite{fisher1993book}. For a two-parameter space we employed a value of $M^p = 1600$, which is sufficient to provide a good approximation of the estimation experiments in the investigated regimes. 

Furthermore, we applied the resampling technique \cite{liu2000book} throughout the data analysis. Such technique is employed to avoid a common issue that may arise in particle approximations. In particular, for progressively larger $N$ and corresponding lower uncertainty on the parameters, most of the weights $w_i$ will be zero-valued. In this case, the set of particles is no more informative on the experiment, and the quantities evaluated from Eqs. (\ref{eq:est_particles}-\ref{eq:sigma_particles}) fail to provide a good approximation of the exact values. To avoid such issue, a new set of particles is periodically resampled \cite{liu2000book} throughout the estimation experiments. This method consists in checking, at each step of the protocol, the concentration of the weight values. If $1/\sum_{i} w_{i}^{2} < M_{\mathrm{th}}$ (in our case, the choice is $M_{\mathrm{th}} = M^p/2$), a new set of particle positions and weights ($\{ \boldsymbol{y}^{\prime}_{i}\}, w_{i}^{\prime}$) is generated according to the current knowledge on the parameters. More specifically, $M^{p}$ new particle positions are generated by random choice from the current ones $\{ \boldsymbol{y}_{i}\}$ according to the weights $w_i$. Then, the new set of particle positions is perturbed according to a multivariate normal distribution having mean $\boldsymbol{\mu}_i = a \boldsymbol{y}_{i} + (1-a) \hat{\boldsymbol{\phi}}$ and covariance matrix $\tilde{\boldsymbol{\Sigma}} = (1-a^{2}) \boldsymbol{\Sigma}$. Here, $\hat{\boldsymbol{\phi}}$ and $\boldsymbol{\Sigma}$ are the mean and covariance of particles before resampling ($\{ \boldsymbol{y}_{i}\}, w_{i}$) as defined in Eqs. (\ref{eq:est_particles}-\ref{eq:sigma_particles}), and $a$ is a parameter included in the interval $a \in [0,1]$. In our case, we used $a=0.98$ according to Refs. \cite{granade2012njp,valeri2020npj}. Once the new particle positions are determined, the new weights are reset to uniform $w_{i}^{\prime} = 1/M^p$.

\section{Results}

Two-phase estimation is performed by Bayesian updating of the {\it a priori} distribution $\mathcal{A}(\phi_1,\phi_2)$ based on the reconstructed probabilities $P_{ML}(x\vert\phi_1,\phi_2)$. In the asymptotic limit of large samples $N\gg 1$, the only relevant figure to assess the uncertainty of the measurement is the local Fisher information $F(\phi_1,\phi_2)$: usually, in this regime the output posterior $p(\phi_1, \phi_2 \vert \boldsymbol{x})$ is sufficiently narrow that details on the prior distribution become irrelevant -- except for ruling out possible ambiguities occurring however at a large scale. With a smaller sample, $N\leqslant 200$, both $P_{ML}(x\vert\phi_1,\phi_2)$ and $\mathcal{A}(\phi_1,\phi_2)$ are relevant: this regime is the one we will be exploring. 

We first consider an {\it a priori} distribution in the Gaussian form: $\mathcal{A}(\phi_1,\phi_2) \propto \exp \left (-\frac{1}{2}(\phi_1{-} \mu_1, \phi_2{-} \mu_2) \cdot \Gamma^{-1} \cdot( \phi_1{-} \mu_1, \phi_2{-} \mu_2)^ \top \right)$, where $\mu_1$ and $\mu_2$ are the mean values, and $\Gamma$ captures the initial uncertainties and the correlations on the parameters. For equal uncertainties $\Gamma_{1,1}= \Gamma_{2,2}$, the width of the prior distribution is quantified by a single parameter $\sigma = \sqrt{\Gamma_{1,1}}$. In particular, we quantify the correlations by means of the Pearson coefficient \cite{pearson1895vii} $\rho = \Gamma_{1,2}/(\Gamma_{1,1}\Gamma_{2,2})^{1/2}$. Similarly, we can introduce a quantity for the Fisher information $\nu= -F_{1,2}/(F_{1,1}F_{2,2})^{1/2}$, where the additional minus sign here originates from the inversion of $F$ in order to obtain a covariance matrix. 

\begin{figure}[ht!]
    \centering
    \includegraphics[width = 0.49 \textwidth, keepaspectratio]{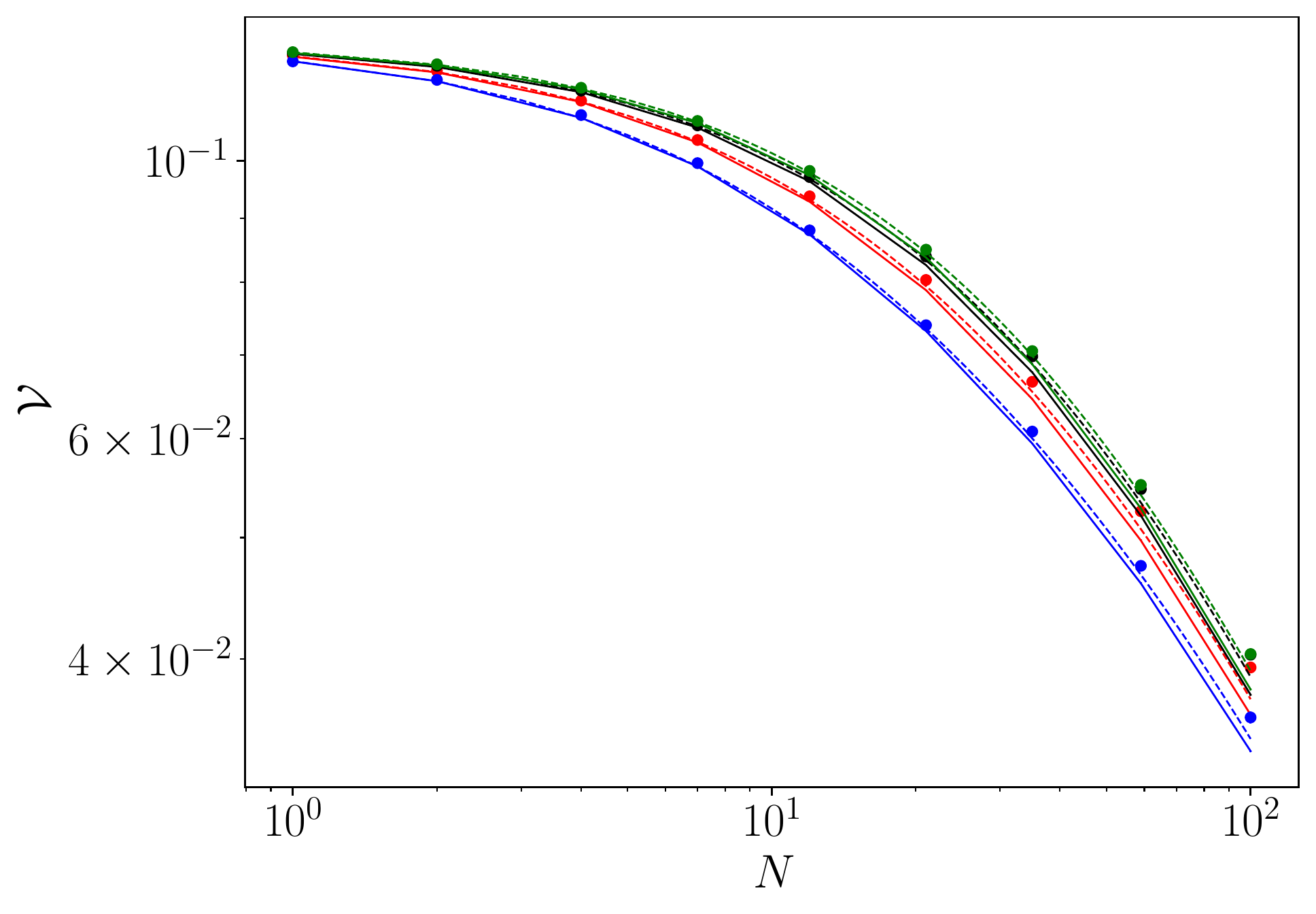}
    \caption{Analysis of the total variance $\mathcal{V}$ as a function of the number of measured events $N$, and comparison with the expected bounds. Plots correspond to $\mu_1, \mu_2 = (1.1, 2.0)$, $\sigma = 0.25$ and different values of correlations $\rho$ in the prior. Red plots: $\rho = 0.0$. Black plots: $\rho = 0.25$. Blue plots: $\rho = -0.25$. Green plots: $\rho = 0.4$. Solid lines: Ziv-Zakai bound. Dashed lines: Van Trees bound. Points: estimates based on experimental data. Error bars on the total variances ($\mathcal{V}$) estimated from the experimental data are obtained as the standard error on $K = 300$ repeated runs, and are smaller than the point size in the plot.}
    \label{fig:results1}
\end{figure}

The relevant figure in our case is the total variance ${\mathcal{V}=\text{Tr}\left[\Sigma\right]}$. Fig.~\ref{fig:results1} shows the scaling of this variance on the phases as a function of $N$. The points, corresponding to our estimates based on the experimental probabilities, are efficiently captured by both the ZZ and the VT bounds, with the former becoming marginally looser for increasing $N$.
Since we have to consider two parameters, the minimal variance at the ZZ bound $\mathcal{V}_{ZZ}$ writes $\mathcal{V}_{ZZ}=\mathcal{Z}({\bf u}_1)+\mathcal{Z}({\bf u}_2)$, where ${\bf u}_1$ and ${\bf u}_2$ are orthogonal unit vectors.
We found that the tighter bound on $\mathcal{V}$ corresponds to choosing ${\bf u}_1$ and ${\bf u}_2$ as the eigenvectors of the covariance matrix $\Sigma$, since these are associated to independent parameters.

\begin{figure*}[ht!]
    \centering
    \includegraphics[width = 0.99 \textwidth, keepaspectratio]{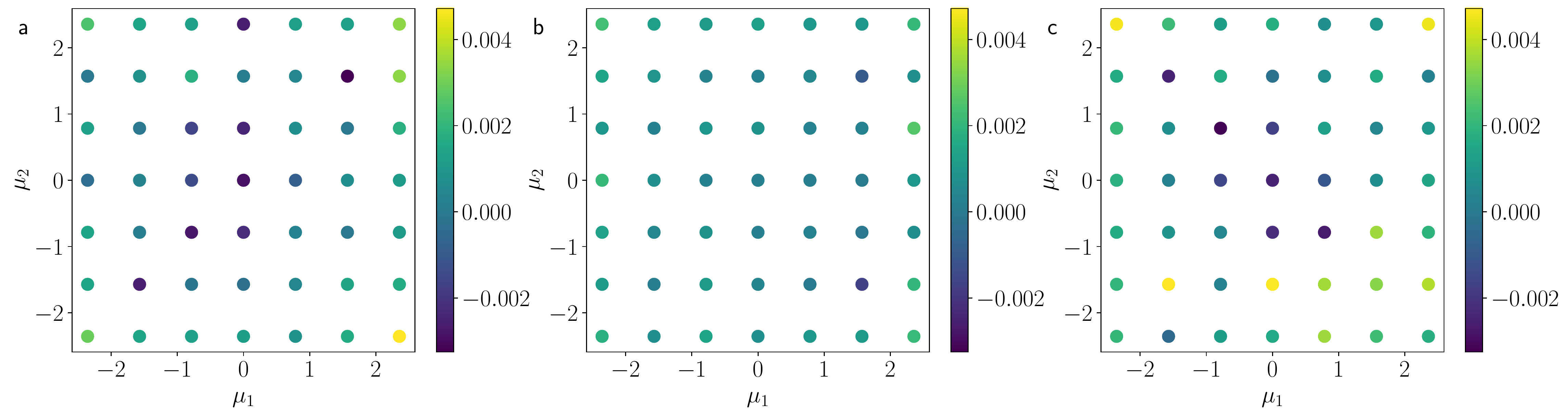}
\caption{Relative difference $(\mathcal{V} - \mathcal{V}_{ZZ})/\mathcal{V}_{ZZ}$ between the total variance $\mathcal{V}$ and the expected Ziv-Zakai bound $\mathcal{V}_{ZZ}$ for $N=1$ by varying the correlation $\rho$ in the prior. Plots are generated for different values of $\mu_1, \mu_2$, and $\sigma = 0.2$. Error bars on the total variances ($\mathcal{V}$) estimated from the experimental data are obtained as the standard error on $K = 300$ repeated runs, and are $<1\%$ (relative error). \textbf{a}, \textbf{$\rho = \nu$}. \textbf{b}, $\rho = 0$. \textbf{c}, $\rho = -1/\nu$.}
    \label{fig:results2}
\end{figure*}

\begin{figure}[ht!]
    \centering
    \includegraphics[width = 0.49\textwidth, keepaspectratio]{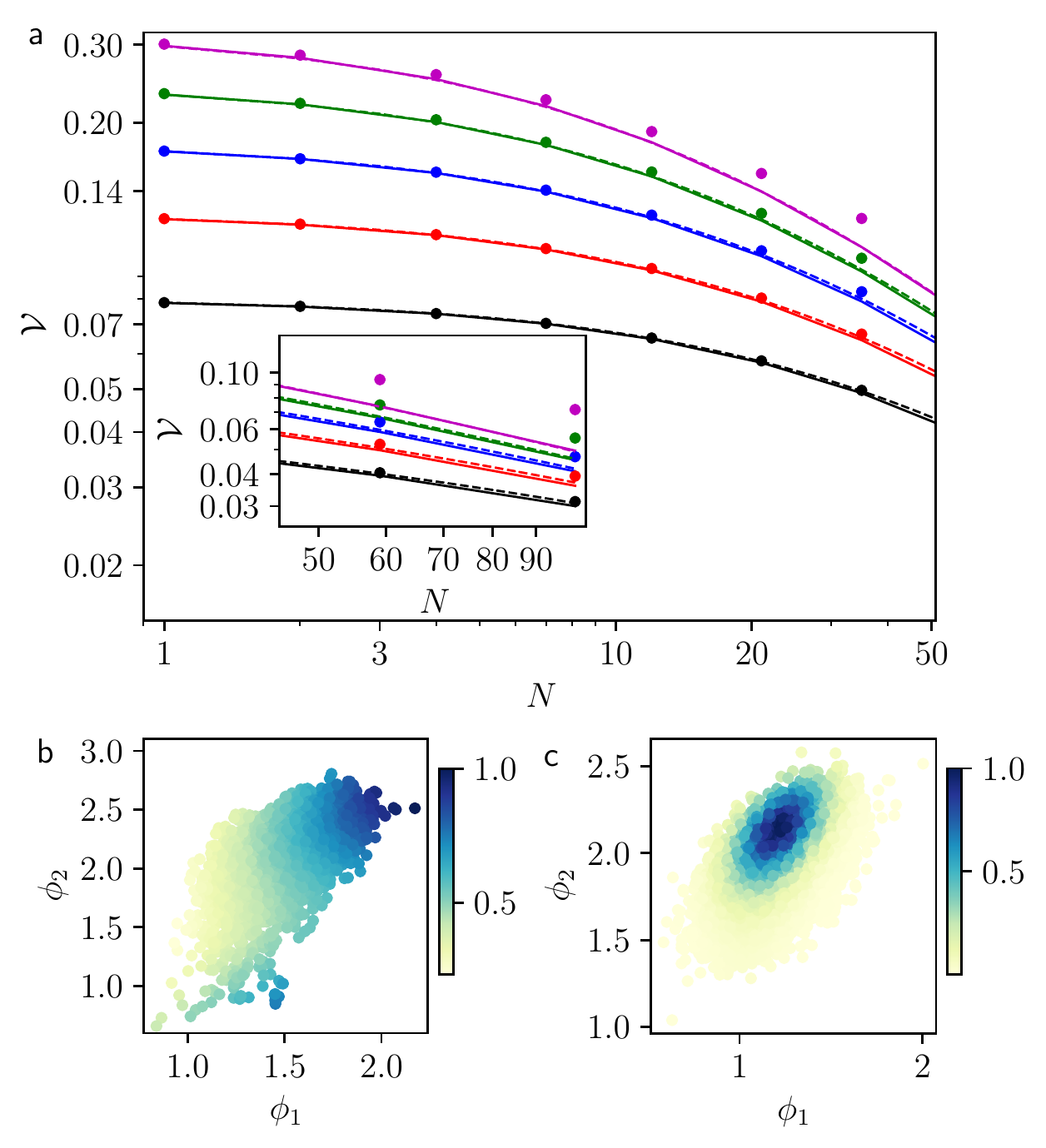}
\caption{Comparison between bounds and experimental estimates for different values of the prior width $\sigma$. \textbf{a}, Total variance $\mathcal{V}$ for $\mu_1, \mu_2 = (1.1, 2.0)$ and $\rho = 0$. Magenta plots: $\sigma = 0.4$. Green plots: $\sigma = 0.35$. Blue plots: $\sigma = 0.3$. Red plots: $\sigma = 0.25$. Black plots: $\sigma = 0.2$. Solid lines: Ziv-Zakai bound. Dashed lines: Van Trees bound. Points: estimates based on experimental data. Error bars on the total variances ($\mathcal{V}$) estimated from the experimental data are obtained as the standard error on $K = 300$ repeated runs, and are smaller than the point size in the plot. \textbf{b}-\textbf{c}, Posterior distributions after an estimation experiment. Here, colors represent the weights $w_i$ for the different particles $\boldsymbol{y}_{i}$. \textbf{b}, Posterior after $N = 100$ events, starting from a prior distribution with $\mu_1, \mu_2 = (1.1, 2.0)$, $\rho = 0$ and $\sigma = 0.4$. \textbf{c}, Posterior after $N = 100$ events, starting from a prior distribution with $\mu_1, \mu_2 = (1.1, 2.0)$, $\rho = 0$ and $\sigma = 0.2$.}
    \label{fig:results3}
\end{figure}

A first case in this plot considers the scenario of a symmetric $\mathcal{A}(\phi_1,\phi_2)$, $\Gamma_{1,1}=\Gamma_{2,2}$, presenting no correlation between the two parameters, $\rho=0$. Other possibilities are illustrated in Fig.~\ref{fig:results1}, for the same widths, but different values of $\rho$: the bounds are still able to capture the variance efficiently, but it must be remarked that the uncertainties themselves are varied. In particular, it can be verified that the minimal uncertainties are achieved when the correlations in the prior distribution match those of the Fisher information: $\rho=\nu$, {\it i.e.} in the absence of competing symmetries in the updated probability. The particular case of Fig.~\ref{fig:results1} exemplifies a general behaviour. The grids in Fig.~\ref{fig:results2} show that the relative variation between the assessed variance at $N=1$ and the corresponding ZZ bound remains below $1\%$, regardless the initial correlation $\rho$. These results also provide  upper limits for the behaviour of the VT bound, which appears to be tighter (see Fig.~\ref{fig:results1}).

The second key parameter in the {\it a priori} is its width, which is varied in the plots of Fig.~\ref{fig:results3}a. Here we show the variance as a function of $N$, for different values of $\Gamma_{1,1}$. As expected, the narrower the prior distribution, the more precise the estimate is.  Crucially, both the ZZ and the VT bounds become less strict for wider prior distributions. The origin of this behaviour can be traced in the form of the posterior distributions, shown in Fig.~\ref{fig:results3}b-c. When $\mathcal{A}(\phi_1,\phi_2)$ is narrower, it eventually produces a well-behaved {\it a posteriori} distribution, with a single peak. In the opposite limit of a wide {\it a priori} distribution, the final distribution is affected by ambiguities in the outcome probabilities, on which $\mathcal{A}(\phi_1,\phi_2)$ does not provide enough information for a discrimination. 
This emphasises the impact of ambiguities in the output distributions. In fact, different phase settings may lead to identical outcomes. The settings can only be discriminated if the {\it a priori} distribution is sufficiently narrow, regardless of the number of copies.  The bounds, on the other hand, do not account for this increased complexity in the posterior. 

\begin{figure}[ht!]
    \centering
    \includegraphics[width = 0.49 \textwidth, keepaspectratio]{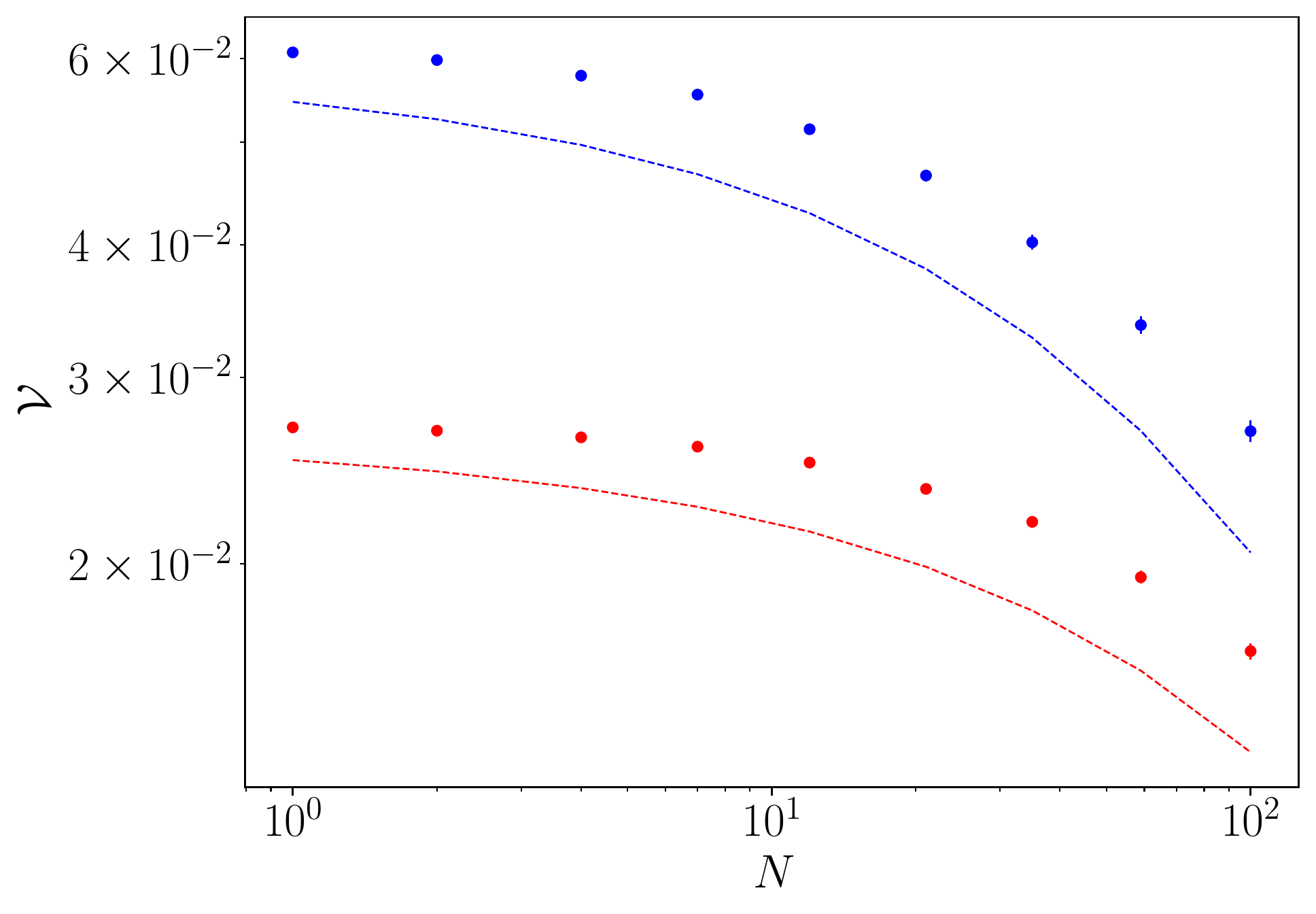}
\caption{Analysis of the total variance $\mathcal{V}$ with a non-derivable prior as a function of the number of events $N$. The chosen prior distribution is a 2D Heavyside rectangular function, with equal width $\Delta$ on each direction. Solid lines: Ziv-Zakai bound. Points: estimates based on experimental data. Error bars on the total variances ($\mathcal{V}$) estimated from the experimental data are obtained as the standard error on $K = 300$ repeated runs, and are smaller than the point size in the plot. Blue plots: $\Delta = 0.6$. Red plots: $\Delta = 0.4$. For all datasets, $\mu_1, \mu_2 = (1.1, 2.0)$. }
    \label{fig:results4}
\end{figure}

In general, the VT bound, which is considerably simpler to calculate, performs better than the more sophisticated ZZ bound. However, this latter presents the advantage of working also for non-derivable {\it a priori} distributions, as for the case presented in Fig.~\ref{fig:results4}. We show the bounds as a function of $N$ when the {\it a priori} distribution is now in the form of a 2D Heaviside rectangle of width $\Delta$. The bounds are not as accurate as in the analytical cases of regular {\it a priori} functions, providing only an estimate for the magnitude of the variance.   

\section{Discussion}

Parameter estimation using quantum resources promises the development of a novel generation of sensors with enhanced precision capabilities. Besides the technological advances, this development must be inevitably accompanied by appropriate data analysis techniques to fully exploit the sensors precision. Notable frameworks to be addressed require going beyond local estimation and asymptotic scenarios, thus dealing with limited resources and different prior knowledge. In this general framework, suitable bounds need to be identified to properly address the performance of quantum estimation experiments.

Here, we have investigated Bayesian multiphase estimation with different prior knowledge in integrated multiarm interferometers. The latter represents a promising platform for quantum sensing, and at the same time can be used as a testbench to develop proper techniques for multiparameter estimation. Within this system, we have investigated the interplay between the natural correlation on the unknown parameters, related to the Fisher information matrix, and the correlation encoded in the prior distribution. Such scenario has been assessed by using two different bounds for Bayesian estimation, that can naturally take into account the availability of different prior knowledge. We observe that both Van Trees and Ziv-Zakai bounds provide a strict estimate of the estimation error for regular and narrow prior distribution. This is observed independently of the relative orientation of the prior distribution correlation with respect to the system Fisher matrix. For larger priors, the bounds become less accurate in the presence of ambiguities in the system output probabilities, which affect the estimation precision. Finally, when the prior distribution is not regular the Ziv-Zakai bound still provides a reasonable, albeit not strict, estimate of the process features.

The present analysis provides a detailed insight on the role of the different quantities underlying multiparameter estimation, in the general regime with limited resources and different prior knowledge. Such results are expected to be of relevance for the development of quantum sensors, and can stimulate further investigation in defining a general and comprehensive framework for multiparameter quantum estimation.

\section*{Acknowledgments}
This work is supported by the European Union's Horizon 2020 research and innovation programme under the PHOQUSING project GA no. 899544, the STORMYTUNE project GA no. 899587, by the Amaldi Research Center funded by the  Ministero dell'Istruzione dell'Universit\`a e della Ricerca (Ministry of Education, University and Research) program ``Dipartimento di Eccellenza'' (CUP:B81I18001170001) and by MIUR (Ministero dell’Istruzione, dell’Università e della Ricerca) via project PRIN 2017 “Taming complexity via QUantum Strategies a Hybrid Integrated Photonic approach” (QUSHIP) Id. 2017SRNBRK. N.S. acknowledges funding from Sapienza Universit\`a via Bando Ricerca 2018: Progetti di Ricerca Piccoli, project "Multiphase estimation in multiarm interferometers". M.B. acknowledges discussion with Marco G. Genoni.

\providecommand{\noopsort}[1]{}\providecommand{\singleletter}[1]{#1}%

\end{document}